\documentclass[sigconf, natbib = false]{acmart}

\usepackage[utf8]{inputenc}
\usepackage{enumitem}
\usepackage{array}
\usepackage{multirow}
\usepackage{multicol}
\usepackage{makecell}
\usepackage{mathtools}
\usepackage[labelfont={sc}]{caption}
\usepackage[capitalize]{cleveref}
\usepackage{todonotes}
\usepackage{moresize}
\usepackage[algoruled,vlined,english,linesnumbered]{algorithm2e}

\RequirePackage[
  style=acmnumeric, 
  backend=biber,
  natbib=true,
  giveninits=true,
  ]{biblatex}
\renewbibmacro{in:}{}
\AtEveryBibitem{\clearfield{month}} 
\AtEveryBibitem{\clearfield{day}}
\AtEveryBibitem{\clearfield{isbn}}
\AtEveryBibitem{\clearfield{issn}}
\AtEveryBibitem{\clearlist{language}}
\renewbibmacro*{doi+eprint+url}{%
    \printfield{doi}%
    \newunit\newblock%
    \iftoggle{bbx:eprint}{%
        \usebibmacro{eprint}%
    }{}%
    \newunit\newblock%
    \iffieldundef{doi}{%
        \usebibmacro{url+urldate}}%
        {}%
    }
\newcommand{\Q}{\mathbb{Q}}

\newcommand{\R}{\mathbb{R}}
\newcommand{\Z}{\mathbb{Z}}
\newcommand{\N}{\mathbb{N}}

\newcommand{\Fp}{\mathbb{F}_p}

\newcommand{\ps}[1]{[\![ #1 ]\!]}
\newcommand{\ua}{\boldsymbol{\alpha}}
\newcommand{\ub}{\boldsymbol{\beta}}

\newcommand{\val}{\text{\rm val}}
\newcommand{\nuval}{\val_{p,\nu}}

\definecolor{input}{HTML}{303060}
\definecolor{output}{HTML}{804000}
\definecolor{string}{HTML}{A02020}
\definecolor{ring}{HTML}{A020A0}
\definecolor{function}{HTML}{205080} 
\definecolor{constructor}{HTML}{205080}
\definecolor{method}{HTML}{205080}
\definecolor{keyword}{HTML}{008000}
\definecolor{error}{HTML}{B01010}
\definecolor{comment}{HTML}{606060}

\newcommand{\cIn}{{\color{input} \tt In}:}
\newcommand{\cOut}{{\color{red} \tt Out}:}

\def\pFq#1#2#3#4#5{{}_{#1}F_{#2}\left(#3, #4; #5\right)}

\setlist[itemize]{leftmargin=1.5em}

\begin{abstract}
    We report on implementations for algorithms treating algebraic and arithmetic
    properties of hypergeometric functions in the computer algebra system SageMath.
    We treat hypergeometric series over the rational numbers, over finite fields,
    and over the $p$-adics. Among other things, we provide implementations deciding
    algebraicity, computing valuations, and computing minimal polynomials in
    positive characteristic.
\end{abstract}

\setcopyright{none}
\renewcommand\footnotetextcopyrightpermission[1]{} 
\acmConference[ISSAC'26]{ACM Conference}{July 2026}{Oldenburg, Germany}

\title[Hypergeometric Functions in SageMath]{Algebraic and Arithmetic Attributes \\
of Hypergeometric Functions in SageMath}
\author{Xavier Caruso}
\affiliation{
	\institution{CNRS; IMB, Université de Bordeaux}
    \city{Talence}
    \country{France}
	}
\email{xavier@caruso.ovh}
\author{Florian Fürnsinn}
\affiliation{
	\institution{University of Vienna, Faculty of Mathematics}
    \city{Vienna}
    \country{Austria}
	}
\email{florian.fuernsinn@univie.ac.at}

\begin{document}

\SetKwInOut{Input}{input}
\SetKwInOut{Output}{output}
\SetKw{KwOr}{or}
\SetKw{KwAnd}{and}
\SetKw{Continue}{continue}
\SetKw{Pop}{pop}
\SetKw{Append}{append}

\thanks{
    The second named author was funded by a DOC Fellowship (27150) of the
    \href{https://www.oeaw.ac.at/en/}{Austrian Academy of Sciences} at the
    University of Vienna. Further he thanks the French–Austrian project EAGLES
    (ANR-22-CE91-0007 \& FWF grant 
    \href{https://doi.org/10.55776/I6130}{10.55776/I6130}) for financial 
    support.

    The authors thank \href{https://oead.at/en/}{Austria’s Agency for Education
    and Internationalisation (OeAD)} and Campus France for providing funding
    for research stays via WTZ collaboration project/Amadeus project FR02/2024.
}

\maketitle

\section{Introduction}

A \emph{hypergeometric function} with rational \emph{top parameters}
$\ua \coloneqq  (\alpha_1,\ldots,\allowbreak \alpha_n)\in \Q^n$ and \emph{bottom parameters} 
$\ub \coloneqq (\beta_1,\ldots, \beta_m) \in \Q^m$
is defined as the power series 
\begin{equation}
\label{eq:pFq}
\pFq{n}{m}{\ua}{\ub}{x} \coloneqq
 \sum_{k=0}^\infty \frac{(\alpha_1)_k\cdots(\alpha_n)_k}
 {(\beta_1)_k\cdots (\beta_m)_k}\cdot \frac{x^k}{k!}\in \Q\ps{x},
\end{equation}
where $(\gamma)_k \coloneqq \gamma (\gamma+1)\cdots (\gamma+k-1)$
denotes the \emph{rising factorial} or \emph{Pochhammer symbol}.


The hypergeometric function $\pFq{n}{m}{\ua}{\ub}{x}$ is the solution of the differential equation
$\mathcal L(\ua, \ub, x) y(x) = 0$
with, denoting $\vartheta = x \frac{\mathrm d}{\mathrm d x}$,
\[\mathcal L(\ua, \ub, x) \coloneqq \big(x (\vartheta{+}\alpha_1)\cdots (\vartheta{+}\alpha_n) -
 \vartheta (\vartheta{-}\beta_1)\cdots (\vartheta{-}\beta_m)\big).\]
 
A power series $f(x)\in K \ps x$ is called \emph{algebraic}
if there exists a nonzero polynomial $P(x,y)\in K[x,y]$, such that
$f(x, y(x))=0$. Similarly, a power series $f(x)\in \Q \ps x$ is called
\emph{D-finite}, if there exists a nonzero differential operator
$L \in \Q[x]\langle \partial\rangle$, such that $Lf(x)=0$. Moreover,
it is called \emph{globally bounded}, if it has a a positive radius
of convergence and there exist two nonzero integers
$\alpha, \beta\in \Z,$ such that $\beta f(\alpha x) \in \Z\ps x$.

Hypergeometric functions are clearly D-finite, and the sets of parameters for
which they are algebraic or globally bounded are fully classified. Moreover,
their reductions modulo $p$ in $\Fp\ps x$ are algebraic, whenever they are
defined. These properties make them ideal test cases for conjectures about
algebraic and arithmetic properties of D-finite series.
The need to use a software for formulating and checking such conjectures
appeared to the authors when they worked, jointly with Vargas-Montoya, on
Galois groups of reductions modulo primes of D-finite series~\cite{CFV25}.

The present article is the outcome of this observation.
It reports on an open source SageMath package designed to manipulate 
hypergeometric functions in various situations: over the rationals, over
finite fields and over $p$-adic fields. To provide these functionalities
it was necessary to develop new algorithms. They are presented mostly
in~\cite{CF26}.

At the time being, our package is submitted for integration in a future
release of SageMath~\cite{github}. In this paper, we shortly showcase the 
main features of our package.
Our presentation is complemented by an interactive worksheet gathering
all the examples presented below, available at:

\noindent
{\color{magenta}%
\url{https://xavier.caruso.ovh/notebook/hypergeometric-functions/}}

\noindent
This interactive worksheet supports editing the examples and exploring
new cases.

For the complete documentation of our package, we refer to the Section 
\emph{Hypergeometric functions over arbitrary rings} in the reference
manual of SageMath, available online \href{https://doc-pr-41113--sagemath.netlify.app/html/en/reference/functions/sage/functions/hypergeometric_algebraic}{\color{magenta} here}.


\section{Setup}

Hypergeometric functions are implemented in SageMath as elements of a symbolic
ring. 

\vspace{1.5mm}

{\noindent \small
\begin{tabular}{rl}

\cIn & \verb?f = hypergeometric([1/3, 2/3], [1/2], x)? \\
\cIn & \verb?f? \\
\cOut &  \verb?hypergeometric((1/3, 2/3), (1/2,), x)? \\
\cIn & \verb?f.parent()? \\
\cOut & \verb?Symbolic Ring?

\end{tabular}}

\vspace{1.5mm}

\noindent
We propose an implementation of algebraic and arithmetic properties of
them, accessed by creating hypergeometric functions where the argument is an
element of (the fraction field of) a polynomial ring or a power series ring.

\vspace{1.5mm}

{\noindent \small
\begin{tabular}{rl}

\cIn & \verb?S.<x> = QQ[]? \\
\cIn & \verb?f = hypergeometric([1/3, 2/3], [1/2], x)? \\
\cIn & \verb?f? \\
\cOut &  \verb?hypergeometric((1/3, 2/3), (1/2,), x)? \\
\cIn & \verb?f.parent()? \\
\cOut & \verb?Hypergeometric functions in x over Rational Field?

\end{tabular}}

\vspace{1.5mm}

We emphasize that our package allows for quite general parameters,
even permitting to have nonnegative integers as bottom parameters 
as soon as they are compensated by an appropriate top parameter.
Compare:

\vspace{1.5mm}

{\noindent \small
\begin{tabular}{rl}

\cIn & \verb?hypergeometric([-1], [-2], x)? \\
\cOut & \verb?hypergeometric((-1,), (-2,), x)? \\
\cIn & \verb?hypergeometric([-2], [-1], x)? \\
\cOut & \verb?ValueError: the parameters ((-2,), (-1,))? \\
      & \verb? do not define a hypergeometric function?
\end{tabular}}

\vspace{1mm}

%
%
%
%
%

\noindent
The main functions we will use throughout the following demonstration are
$f(x)$, Christol's example of a hypergeometric function that is not known 
to be a diagonal; $g(x)$, where $g(16x^2)$ is the generating
function of Gessel excursions, and the 
somewhat obscure function $h(x)$ that illustrates many phenomena.

\vspace{1.5mm}

{\noindent \small
\begin{tabular}{rl}

\cIn & \verb?f = hypergeometric([1/9, 4/9, 5/9], [1/3, 1], x)? \\
\cIn & \verb?g = hypergeometric([1/2, 5/6, 1], [5/3, 2], x)? \\
\cIn & \verb?h = hypergeometric([1/5, 1/5, 1/5, 1/5],? \\
     & \verb? [1/3, 3^10/5 - 1], x)?

\end{tabular}}

\vspace{1.5mm}

\noindent

\section{Hypergeometric functions over $\Q$} \label{sec:Q}

\smallskip

\paragraph{Global Boundedness}

Globally bounded hypergeometric functions have been fully classified
by Christol~\cite{Chr86}. We implement his criterion in the
method \verb?is_globally_bounded()?.

\vspace{1.5mm}

{\noindent \small
\begin{tabular}{rl}

\cIn & \verb?f.is_globally_bounded()? \\
\cOut & \verb?True? \\
\cIn & \verb?h.is_globally_bounded()? \\
\cOut & \verb?False? \\

\end{tabular}}

\vspace{1.5mm}

\noindent
Christol's elegant criterion, essentially depends
on the relative positions of the decimal parts $\{\Delta \alpha_i\}$ and
$\{\Delta \beta_i\}$, for $\Delta \in \Z/d\Z$, where $d$ is the least
common denominator of the parameters.

\paragraph{Algebraicity}
Similar in spirit, all algebraic hypergeometric functions have been classified.
We implemented the corresponding criterion in the method \verb?is_algebraic()?.

\vspace{1.5mm}

{\noindent \small
\begin{tabular}{rl}

\cIn & \verb?f.is_algebraic()? \\
\cOut & \verb?False? \\
\cIn & \verb?g.is_algebraic()? \\
\cOut & \verb?True? \\

\end{tabular}}

\vspace{1.5mm}

\noindent
In case there are no integer differences between top and bottom parameters
Christol~\cite{Chr86}, and Beukers and Heckman~\cite{BH89} provided the classification,
a criterion, very similar to the classification of global boundedness,
depending on relative positions of decimal parts.
In case of integer differences, the criterion was extended by the second author
and Yurkevich~\cite{FY24}. 

\paragraph{Primes With Good Reductions}

We say that a series $s(x) \in \mathbb Q\ps x$ has good reduction at $p$,
if all its coefficients have denominator coprime with $p$. 
For large enough prime numbers, this property only depends on the congruence 
class of $p$ modulo $d$.
The method \verb?good_reduction_primes()? computes this
and outputs a set of prime numbers depending on congruence classes (implemented
in SageMath by the first author \cite{githubprimes}).

\vspace{1.5mm}

{\noindent \small
\begin{tabular}{rl}

\cIn & \verb?g.good_reduction_primes()? \\
\cOut & \verb?Set of all prime numbers with 2 excluded:?\\
      & \verb? 3, 5, 7, 11, ...?\\
\cIn & \verb?h.good_reduction_primes()? \\
\cOut & \verb?Set of prime numbers congruent to 1, 8, 11 modulo 15? \\
 & \verb? with 17, 167, 677, 857, ..., 29327, 29387 included?\\
 & \verb? and 23, 83, 113, 173, ..., 58913, 58943 excluded:? \\
 & \verb? 11, 17, 31, 41, ...?
\end{tabular}}

\vspace{1.5mm}

\noindent
Similar to the techniques used for the classification of globally bounded
hypergeometric functions, the authors showed in \cite{CFV25} and \cite[\S~3.1]{CF26}
that for large enough prime numbers it only depends on their congruence class
modulo $d$, whether the series has nonnegative $p$-adic valuation, \emph{i.e.}, can 
be reduced modulo $p$. Our implementation tests for small prime numbers $p$,
whether the $p$-adic valuation is positive, using the method
\verb?valuation()?, showcased in Section~\ref{sec:val}. This is done for all primes
smaller than the bound, after
which regularity is ensured, and additionally for one prime number for each
congruence class larger than this bound. We remark that alternative methods
closely related to Christol's work were detailed in \cite[\S~3.1]{CFV25}, which are
not implemented.

\section{Hypergeometric functions over $\Fp$} \label{sec:Fp}

When a hypergeometric function $h(x)$ has good reduction at a prime $p$, 
we can form $h(x) \bmod p \in \Fp\ps x$ and study its properties.
It is known for example that the latter is always 
algebraic~\cite{Chr86a, Var21, Var24, CFV25, CF26}.
The main ingredients beyond this result are \emph{section operators}, 
which allow to build what we call \emph{Dwork relations} (a writing of 
hypergeometric function as a polynomial linear combination of $p$-th powers 
of other ones) and eventually to find annihilating
polynomials. All these steps are implemented in our package.

\noindent
\paragraph{Sections}
For an integer $r$, we define the $r$-th section operator
\[ \begin{array}{rcl}
    \Fp\ps x &\longrightarrow& \Fp\ps x \smallskip \\
    \sum_{k=0}^\infty a_k x^k &\mapsto& \sum_{k=0}^\infty a_{kp+r} x^k
\end{array}\]
In \cite[\S 3.2]{CF26}, it was shown that sections
of hypergeometric functions can always be written as a product of a
monomial and another hypergeometric function.
Those can be computed using the method \verb?section()?.

\vspace{1.5mm}

{\noindent \small
\begin{tabular}{rl}

\cIn & \verb?f19 = f % 19? \\
\cIn & \verb?f19.section(0)? \\
\cOut & \verb?hypergeometric((1/9, 4/9, 5/9), (1/3, 1), x)? \\
\cIn & \verb?f19.section(1)? \\
\cOut & \verb?14*hypergeometric((1/9, 4/9, 5/9), (1/3, 1), x)? \\
\cIn & \verb?f19.section(8)? \\
\cOut & \verb?5*hypergeometric((4/9, 5/9, 10/9), (1, 4/3), x)? \\
\cIn & \verb?f19.section(10)? \\
\cOut & \verb?0? \\
\end{tabular}}

\vspace{1.5mm}

\noindent
\paragraph{Dwork Relations}

From what precedes, one can write a hypergeometric functions as a 
$\Fp[x]$-linear
combination of $p$-th powers of other hypergeometric functions. We
call these relations \emph{Dwork relations} and they are implemented
in the method \verb?dwork_relation()?. 
Here the output is a dictionary, where hypergeometric
functions are assigned polynomial coefficients: the sum of the $p$-th power
of the keys, weighted by the assigned values gives the original hypergeometric
function.

\vspace{1.5mm}

{\noindent \small
\begin{tabular}{rl}

\cIn & \verb?f19.dwork_relation()? \\
\cOut & \verb?{hypergeometric((1/9, 4/9, 5/9), (1/3, 1), x):? \\
      & \verb? 8*x^2 + 14*x + 1,?\\
      & \verb? hypergeometric((4/9, 5/9, 10/9), (1, 4/3), x):? \\
      & \verb? 5*x^8 + 15*x^7}?
\end{tabular}}

\vspace{1.5mm}

\paragraph{Annihilating Polynomials}
An annihilating polynomial of a reduction of a hypergeometric function
can be computed with the method \verb?annihilating_ore_polynomial()?,
which outputs a polynomial in the Frobenius: to view it as an actual
polynomial, one should replace $\verb?Frob?^n$ by $X^{p^n}$.

\vspace{1.5mm}

{\noindent \small
\begin{tabular}{rl}

\cIn & \verb?f19.annihilating_ore_polynomial()? \\
\cOut & \verb?(18*x^76 + 13*x^57 + 6*x^38 + 17*x^19 + 12)*Frob^2 +? \\
      & \verb? (12*x^38 + 11*x^32 + ... + 18*x^12 + 7)*Frob +? \\
      & \verb? x^30 + 16*x^29 + ... + 6*x^13 + x^12?

\end{tabular}}

\vspace{1.5mm}

\noindent
An algorithm performing this computation, based on iteratively
computing Dwork relations that lead to a system of equations between
finitely many hypergeometric functions, was described in \cite[\S 3.3]{CFV25}.

This algorithm is implemented essentially.
We warn the user that the current implementation relies on a simplified version
of the algorithm, for which it is not proven that the computed system only 
involves finitely many hypergeometric series, so it might happen that
computations do not terminate. However, we believe that also the current
implementation can be shown to always terminate.

The polynomials obtained this way are never irreducible. By their nature
as linearized polynomials, they are always divisible by $h(x)$. However,
in general, they will factor even further.

\paragraph{Congruences Modulo Primes}
It is possible that two hypergeometric functions with different
set of parameters leads to series which are congruent modulo $p$,
as showcased in the code example below.

\vspace{1.5mm}

{\noindent \small
\begin{tabular}{rl}

\cIn & \verb?T.<y> = GF(13)[]? \\
\cIn & \verb?h1 = hypergeometric([1/12, 1/4], [1/2], y)? \\
\cIn & \verb?h2 = hypergeometric([1/12, 1/6], [1/3], y)? \\
\cIn & \verb?h1.power_series(1000) - h2.power_series(1000)? \\
\cOut & \verb?O(y^1000)?

\end{tabular}}

\vspace{1.5mm}

\noindent
The method \verb?is_equal_as_series()? checks when this happens.

\vspace{1.5mm}

{\noindent \small
\begin{tabular}{rl}

\cIn & \verb?h1.is_equal_as_series(h2)? \\
\cOut & \verb?True?

\end{tabular}}

\vspace{1.5mm}

\noindent
We sketch an algorithm to check when this holds.
Our method relies on the following basic observation: two series are
congruent modulo $p$ if and only if all their $r$-th sections
for
$0 \leq r < p$ are congruent modulo $p$.
Besides, in our case of interest, we can compute the sections of an
hypergeometric series, which we have seen to be constant multiples of 
hypergeometric functions themselves. We can then proceed recursively.
Reusing the arguments we used for \verb?annihilating_polynomial()?, we
conclude that we will only encounter finitely many hypergeometric
series while proceeding.

To implement this strategy we rely on two auxiliary
sets for recording the progress of the algorithm, namely
\begin{itemize}
\item the set $Q$ (for ``queued'') of pairs $(f_1(x), f_2(x))$ whose
congruence modulo $p$ still needs to be checked, and
\item the set $C$ (for ``checked'') of pairs $(f_1(x), f_2(x))$ whose 
congruence modulo $p$ has already been checked.
\end{itemize}

\noindent
At first glance, we only transfer the problem
to deciding whether a finite number of pairs of hypergeometric
functions have the same reduction. However, by checking that all sections
of a pair of such functions have the same constant term, we
check that the pair of functions agree up to order $x^{p-1}$. Thus, 
by only checking equality of \emph{constant terms,} we can decide equality: 
if for any pair encountered the constant terms do not agree, the
reductions of the two hypergeometric functions are distinct, otherwise
we iterative conclude equality up to an arbitrary order of precision,
\emph{i.e.}, the reductions coincide.

\begin{algorithm}[t]

  \Input{$h_1(x), h_2(x), p$}
  \Output{whether $h_1(x) \equiv h_2(x) \pmod p$}

  $Q \leftarrow \{(h_1(x), h_2(x))\}$\;
  $C \leftarrow \emptyset$\;
  \While{$Q \neq \emptyset$}{
    \Pop a pair $(f_1(x), f_2(x))$ from $Q$\;
    \If{$(f_1(x), f_2(x)) \in C$ \KwOr $(f_2(x), f_1(x)) \in C$}{
      \Continue\;
    }
    \For{$r\gets0$ \KwTo $p{-}1$}{
      for $i = 1, 2$,
      write $\Lambda_r(f_i(x)) \bmod p$ as $a_i x^{e_i} g_i(x)$\hspace{3cm}
      with $a_i \in \Fp$, $e_i \in \N$ and $g_i(x)$ hypergeometric\;
      \If{$a_1 x^{e_1} \neq a_2 x^{e_2}$ in $\Fp[x]$}{
        \Return false\;
      }
      \If{$a_1 x^{e_1} \neq 0$ in $\Fp[x]$}{
        \Append $(g_1(x), g_2(x))$ to $Q$\;
      }
    }
    \Append $(f_1(x), f_2(x))$ to $C$\;
  }
  \Return true\;

  \caption{\texttt{are\_congruent} \label{algo:congruence}}
\end{algorithm}


\paragraph{$p$-curvatures}

For hypergeometric functions with $m= n-1$, we implement the 
$p$-curvature of the
associated hypergeometric differential operator $\mathcal L(\ua, \ub; x)$,
i.e., a matrix representation of linear map $\partial^p$ acting on 
$\Fp[x]\langle \partial \rangle / \mathcal L(\ua, \ub; x)\Fp[x]\langle \partial \rangle$.

It can be accessed by the method \verb?p_curvature()?. Its corank determines
the $\Fp(x^p)$ dimension of solutions of the differential
operator $\mathcal L(\ua, \ub; x)$ in $\Fp(x)$.

The corank of the $p$-curvature for a given
hypergeometric varies in $p$ uniformly~\cite[Prop. 3.1.20]{CFV25} for large enough primes $p$.
The corresponding congruence classes
and exceptions are implemented via the method \verb?p_curvature_coranks()? for 
hypergeometric functions defined over $\Q$.

\vspace{1.5mm}

{\noindent \small
\begin{tabular}{rl}

\cIn & \verb?f5 = f % 5? \\
\cIn & \verb?f5.p_curvature()? \\
\cOut & \verb?[              0 2/(x^5 + 4*x^4) 1/(x^4 + 4*x^3)]? \\
      & \verb?[              0               0               0]? \\ 
      & \verb?[              0               0               0]? \\
\cIn & \verb?f.p_curvature_coranks()? \\
\cOut & \verb?{1: Empty set of prime numbers,? \\
      & \verb? 2: Set of all prime numbers with 3 excluded:? \\
      & \verb?    2, 5, 7, 11, ...,? \\
      & \verb? 3: Empty set of prime numbers}?
\end{tabular}}

\section{Hypergeometric functions over $\Q_p$} \label{sec:val}

Last, we deal with hypergeometric functions with rational parameters
defined over the field of $p$-adic numbers $\Q_p$.
For $x \in \Q_p$, we let $\val_p(x)$ denote its $p$-adic valuation and
we let $\Vert x \Vert_p = p^{-\val_p(x)}$ be its $p$-adic norm.

\paragraph{Radius of Convergence}

Similar to the complex case,
the $p$-adic radius of convergence of a series $s(x) = \sum a_k x^k 
\in \Q_p\ps x$ is defined as

\vspace{1.5mm}

\noindent\hfill%
$\liminf_{k \to \infty} \Vert a_k \Vert_p^{-1/k}.$\hfill\null

\vspace{1.5mm}

\noindent
When $\Vert a \Vert_p$ (with $a \in \Q_p)$ is less than
this critical value, the series $s(a)$ converges in $\Q_p$.
The method \verb?log_radius_of_convergence()? computes the logarithm
in base $p$ of the $p$-adic radius of convergence of a hypergeometric
series.

\vspace{1.5mm}

{\noindent \small
\begin{tabular}{rl}

\cIn & \verb?hp5 = h.change_ring(Qp(5))?\\
\cIn & \verb?hp5.log_radius_of_convergence()? \\
\cOut & \verb?-7/2?\\

\end{tabular}}

\vspace{1.5mm}

\noindent
The algorithm for computing the logarithmic $p$-adic radius of convergence of 
$\pFq n m {\ua}{\ub}{x}$ closely follows the discussion of \cite[\S 2]{CF26}.
We first partition $\ua = \ua' \sqcup \ua''$, $\ub = \ub' \sqcup \ub''$,
where $\ua', \ub'$ contain precisely the $p$-adic integers among the parameters.
Then, assuming that $\pFq n m {\ua}{\ub}{x}$ is not a polynomial, 
its logarithmic radius of convergence is given by the explicit formula
$$\frac{n'{-}m'{-}1}{p{-}1} + \sum_{\alpha \in \ua''} \val_p(\alpha) - \sum_{\beta \in \ub''} \val_p(\beta),$$
where $n'$ and $m'$ are the cardinalities of $\ua'$ and $\ub'$ respectively.
On the contrary, when $\pFq n m {\ua}{\ub}{x}$ is a polynomial, the
logarithm radius of convergence is of course infinite.

\paragraph{Valuations}
For $\nu\in \Q$, we call
$$\textstyle 
\nuval (s(x)) \coloneq \min_{k\geq 0} \, \val_p(a_k) + \nu k$$
the \emph{$\nu$-drifted $p$-adic valuation} of the series $s(x)$.
For $\nu=0$, it clearly coincides with the $p$-adic Gauss valuation of $s(x)$,
and for arbitrary $\nu$, it can be interpreted as the $p$-adic valuation of
$s(x)$ on a disk of $p$-adic radius $p^\nu$ centered at $0$. In particular,
it is $-\infty$ when $\nu$ is greater than the logarithmic $p$-adic radius
of convergence.

The method
\verb?valuation()? computes the $p$-adic valuation of a hypergeometric
function, and passing a parameter $\nu$, it computes the $\nu$-drifted
$p$-adic valuation. Additionally one can also pass the option
\verb?position=True?, to also output the minimal index $k$, for which the
valuation is attained for the coefficient in $x^k$.

\vspace{1.5mm}

{\noindent \small
\begin{tabular}{rl}

\cIn & \verb?fp5 = f.change_ring(Qp(5))? \\
\cIn & \verb?fp5.valuation()? \\
\cOut & \verb?0? \\
\cIn & \verb?hp3 = h.change_ring(Qp(3))?\\
\cIn & \verb?hp3.valuation(position=True)? \\
\cOut & \verb?(-4, 2)?\\
\cIn & \verb?hp5.valuation()?\\
\cOut & \verb?-Infinity?\\
\cIn & \verb?hp5.valuation(-7/2)?\\
\cOut & \verb?0?

\end{tabular}}

\vspace{1.5mm}

\noindent
The authors described in \cite[\S 2.2]{CF26} an algorithm how to compute
the $\nu$-drifted $p$-adic valuations of hypergeometric series.
It relies on a recursion over the tropical semi-ring and the Floyd-Warshall
algorithm to compute the weak transitive closure of a tropical matrix.
Keeping track of the minimal index $k$, for which the valuation is attained for
the coefficient $h_k$ is easily possible, as explained
in~\cite[Rem.~2.6]{CF26}.

\paragraph{$p$-adic Evaluations}

When $a$ is a $p$-adic number with norm less than the radius of convergence,
the value $h(a)$ makes sense.
Our package allows to compute it using the following obvious syntax.

\vspace{1.5mm}

{\noindent \small
\begin{tabular}{rl}

\cIn & \verb?fp5(5)? \\
\cOut & \verb?1 + 3*5^2 + 5^4 + ... + O(5^20)? \\
\cIn & \verb?hp3(1/3)? \\
\cOut & \verb?3^-5 + 2*3^-1 + 1 + 2*3 + ... + O(3^13)? \\

\end{tabular}}

\vspace{1.5mm}

\noindent
The implemented algorithm is outlined in~\cite[\S 2.4]{CF26}: given a $p$-adic
number $a\in \Q_p$, with $\val_p(a)=\nu$ within the radius of convergence,
and a precision $N$, we first compute a bound $K$, such than
$\val_p(h_ka^k)>N$ for all $k>K$. Thus $h(a)=\sum_{k=0}^Kh_ka^k+O(p^N)$.
The computation of $K$ depends on the choice of a parameter between $\nu$ 
and the radius of convergence; the heuristics of our choice is also explained 
in~\emph{loc. cit.}

\paragraph{Newton Polygons}

The Newton polygon of a series $\sum_k a_k x^k$ is the convex hull in $\R^2$ of 
the points $(k,v)$ with $v\geq\val_p(h_k)$.

For hypergeometric series, it can be computed with the method
\verb?newton_polygon()?. Passing an argument
$\nu$, with $\nu$ chosen smaller than
the logarithmic $p$-adic radius of convergence, can help to handle cases where the Newton
polygon has an infinite number of slopes. It shrinks the domain of definition
of the hypergeometric series to provide approximations of the Newton polygon. 

\vspace{1.5mm}

{\noindent \small
\begin{tabular}{rl}

\cIn & \verb?hp3.newton_polygon()? \\
\cOut & \verb?ValueError: infinite Newton polygon; try to truncate? \\
      & \verb? it by giving a log radius less than 2? \\
\cIn & \verb?NP = hp3.newton_polygon(7/4)? \\
\cIn & \verb?NP? \\
\cOut & \verb?Infinite Newton polygon with 5 vertices:? \\
      & \verb? (0, 0), (2, -4), (3, -4), (4, -3), (7, 2)?\\
      & \verb? ending by an infinite line of slope 7/4?\\
\cIn & \verb?NP.plot() +?\\
      & \verb? point([(i, hp3[i].valuation()) for i in range(8)])? \\
&
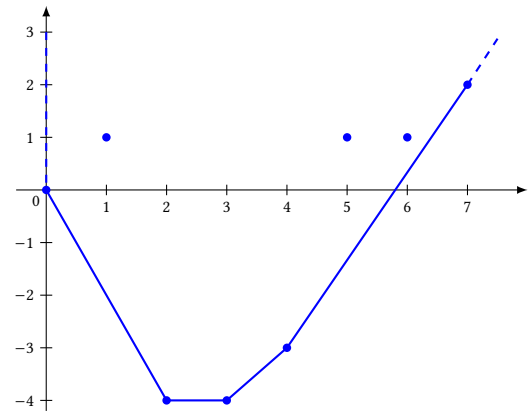
\begin{tikzpicture}[xscale=0.8, yscale=0.7]
\clip (-0.8, -4.3) rectangle (8.2, 3.7);
\draw[-latex] (-0.5, 0)--(8, 0);
\draw[-latex] (0, -4.2)--(0, 3.5);
\foreach \x in {1,...,7} {
  \draw (\x, 0.1)--(\x, -0.1);
  \node[below, scale=0.8] at (\x, -0.1) { $\x$ };
}
\foreach \y in {-4,-3,-2,-1,1,2,3} {
  \draw (0.1, \y)--(-0.1, \y);
  \node[left, scale=0.8] at (-0.1, \y) { $\y$ };
}
\node[below left, scale=0.8] at (0, 0) { $0$ };
\draw[thick, blue] (0, 0)--(2, -4)--(3, -4)--(4, -3)--(7, 2);
\draw[thick, dashed, blue] (0, 3)--(0, 0);
\draw[thick, dashed, blue] (7, 2)--(7.5, 2.875);
\fill[blue] (0, 0) ellipse (0.7mm and 0.8mm);
\fill[blue] (1, 1) ellipse (0.7mm and 0.8mm);
\fill[blue] (2, -4) ellipse (0.7mm and 0.8mm);
\fill[blue] (3, -4) ellipse (0.7mm and 0.8mm);
\fill[blue] (4, -3) ellipse (0.7mm and 0.8mm);
\fill[blue] (5, 1) ellipse (0.7mm and 0.8mm);
\fill[blue] (6, 1) ellipse (0.7mm and 0.8mm);
\fill[blue] (7, 2) ellipse (0.7mm and 0.8mm);
\end{tikzpicture}

\end{tabular}}

\vspace{1.5mm}

\noindent
We implemented the algorithm of \cite[\S 2.3]{CF26}, which is basically
a generalization of the algorithm to compute drifted valuations.

\printbibliography

\end{document}